\documentclass{article}

\usepackage{cite}
\usepackage{amsmath,amssymb,amsfonts}
\usepackage{algorithmic}
\usepackage{textcomp}
\usepackage{xcolor}
\def\BibTeX{{\rm B\kern-.05em{\sc i\kern-.025em b}\kern-.08em
    T\kern-.1667em\lower.7ex\hbox{E}\kern-.125emX}}

\usepackage{rotating}
\usepackage{bm}
\usepackage{multirow}
\usepackage[color]{changebar}
\usepackage{caption}
\usepackage{url}
\usepackage{subfigure}
\usepackage{booktabs}

\usepackage{geometry}
\usepackage{listings}
\begin{document}

\title{Measuring the Quality of B Abstract Machines with ISO/IEC 25010}

\author{Cheng-Hao Cai $~~~~~~~~$ Jing Sun $~~~~~~~~$ Gillian Dobbie \\ School of Computer Science, University of Auckland, Auckland, New Zealand \\ \{chenghao.cai, jing.sun, g.dobbie\}@auckland.ac.nz}

\date{}

\maketitle

\begin{abstract}
The B method has facilitated the development of software by specifying the design of software as abstract machines and formally verifying the correctness of the abstract machines. The quality of B abstract machines can significantly impact the quality of final software products. In this paper, we propose a set of criteria for measuring the quality of B abstract machines based on ISO/IEC 25010, which is one of the latest international standards for evaluating software quality in software engineering. These criteria evaluate abstract machines using a number of general-purpose and domain-independent equations and model checking techniques, so that the quality of abstract machines can be quantified as vectors. The proposed criteria are implemented as a B model quality evaluator, and they are explained and justified using a number of examples.
\end{abstract}

\section{Introduction}

Model checking techniques enable developers to formally verify whether design models of software products satisfy desired properties in their state transition systems \cite{DBLP:books/daglib/0007403}. Given a design model with a number of initial states, operations and desired properties, model checkers such as ProB \cite{DBLP:journals/sttt/LeuschelB08} and NuSMV \cite{DBLP:journals/sttt/CimattiCGR00} can compute a state transition system by iteratively deriving new states from existing states using the operations, and counterexamples violating the desired properties can be reported. However, the counterexamples only indicate single flaws in the state transition system, but may not reflect the overall quality of the design model. If a model checker can measure the overall quality of design models, designers can decide whether and how to further improve the models before actually implementing software products.

Much work has been done on measuring the quality of design models from different aspects. For example, in \cite{DBLP:journals/sqj/NelsonM07}, the semantic quality is used to evaluate the correctness (validity) and completeness of models with respect to real world domains. If a model does not include any representations that contradict with knowledge in a real world domain, then the model is semantically correct. If the model includes all necessary knowledge in the real world domain, then the model is semantically correct. Moreover, in the B-method \cite{DBLP:books/daglib/0015096}, the consistency of a model can be evaluated by verifying whether a model satisfies given specifications. Additionally, simple measurements such as the size of statements and the lines of code have been used to evaluate the quality of code \cite{DBLP:journals/tse/GouesNFW12}. The above studies have provided different criteria of model quality on specific tasks, but have not provided unified measurements of model quality.

The above problem motivates us to develop a set of criteria for measuring the model quality based on ISO/IEC 25010 \cite{ISOIEC25010}, which is a unified international standard for systematically measuring the quality of software products. We develop our criteria based on the B-method \cite{DBLP:books/daglib/0015096}, and they can be extended to other formal model design methods with state transition systems. The \textbf{contributions} of our work include:
\begin{itemize}
\item criteria for measuring functional suitability of formal design models with respect to given functional requirements
\item criteria for measuring security and reliability of formal design models with respect to given invariants
\item criteria for measuring maintainability, performance efficiency and usability of formal design models using model checking techniques
\item the application of the above criteria to the B-method
\item an evaluation for the above criteria on the Volvo cruise controller model.
\end{itemize}

The rest of this paper includes five sections. Section \ref{sec:preliminaries} introduces preliminary knowledge such as ISO/IEC 25010 and the B-method. Section \ref{sec:criteria} introduces the proposed criteria for measuring the quality of formal design models and provides examples to explain the criteria. Section \ref{sec:evaluation} uses the Volvo cruise controller model to evaluate the proposed criteria. Section \ref{sec:related_work} compares the proposed criteria with other existing criteria. Section \ref{sec:conclusion} concludes our work.

\section{Preliminaries}
\label{sec:preliminaries}

In this section, we introduce preliminary knowledge of our work, including ISO/IEC 25010, which is one of the international standards of software quality, and model checking techniques for computing state transition systems.

\subsection{ISO/IEC 25010}
\label{sec:iso_standard}

ISO/IEC 25010 \cite{ISOIEC25010} is one of the latest international standards for systematically measuring the quality of software products. Its latest version was reviewed and confirmed in 2017. The ancestor of ISO/IEC 25010 is ISO/IEC 9126, which has been used to evaluate a wide range of software products in past decades \cite{DBLP:journals/software/JungKC04}. ISO/IEC 25010 includes the following eight characteristics to measure the quality of software products.
\begin{itemize}
\item Functional suitability concerns whether a software product meets functional requirements.
\item Performance efficiency concerns whether a software product can effectively make use of given resources.
\item Compatibility concerns whether a software product can consistently perform its functions while exchanging information and sharing resources with other products.
\item Usability concerns whether users can effectively and efficiently use a software product to complete tasks.
\item Reliability concerns whether a software product can perform desired functions after scheduled to work.
\item Security concerns whether a software product can protect its data and only allows authorised users to access the data.
\item Maintainability concerns whether a software product can be easily modified, repaired and updated by maintainers.
\item Portability concerns whether a software product can be easily used in different environments.
\end{itemize}
The eight characteristics can derive different criteria for different applications. For example, in mobile computing, reliability and security have higher weights than other characteristics \cite{DBLP:conf/apsec/IdriMA13}, and their criteria need to be specifically designed for mobile environments.

\subsection{Model Checking}

Model checking techniques \cite{DBLP:books/daglib/0007403} can derive state transition systems from design models. A design model usually has an initialisation statement that derives one or more initial states and a number of operations that derive state transitions from the initial states. Each operation has a pre-condition describing when the operation can be activated and a post-condition describing the consequence of activating the operation. Model checkers can instantiate the activation of operations using state transitions of the form $ [s,f,t] $, where $ f $ is an operation, $ p $ is an existing state satisfying the pre-condition of $ f $, and $ t $ is a new state satisfying the post-condition of $ f $. All available state transitions form a labelled state transition system of the design model. Moreover, the model checkers can check if the state transition system satisfies desired properties. For example, ProB \cite{DBLP:journals/sttt/LeuschelB08} can check properties written in first-order logic, and NuSMV \cite{DBLP:journals/sttt/CimattiCGR00} can check properties written in linear temporal logic and computational tree logic.

\iffalse

\begin{figure*}
\centering
\includegraphics[width=18cm]{b-model-examples.png}
\caption{Clock models for explaining the criteria of B model quality.}
\label{fig:clock_models}
\end{figure*}

\fi

\begin{figure*}
\centering
\begin{minipage}[t]{0.45\textwidth}
\centering
\begin{lstlisting}[frame=single,basicstyle=\scriptsize,columns=fullflexible]
MACHINE CM1 // Clock Model 1
VARIABLES hour, minute
INVARIANT hour : 0..23 & minute : 0..59
INITIALISATION hour := 0; minute := 0
OPERATIONS
  inc_minute =
    PRE minute < 59
    THEN minute := minute + 1 END;
  inc_hour = PRE minute = 59 & hour < 23
    THEN minute := 0; hour := hour + 1 END;
  next_day =
    PRE minute = 59 & hour = 23
    THEN minute := 0; hour := 0 END END
\end{lstlisting}
\vspace{-3pt}
(a) Clock Model 1 (CM1)
\end{minipage}
~~
\begin{minipage}[t]{0.45\textwidth}
\centering
\begin{lstlisting}[frame=single,basicstyle=\scriptsize,columns=fullflexible]
MACHINE CM2 // Clock Model 2
VARIABLES hour, minute
INVARIANT hour : 0..23 & minute : 0..59
INITIALISATION hour := 0; minute := 0
OPERATIONS
  inc_minute =
    PRE minute < 59
    THEN minute := minute + 1 END;
  inc_hour = PRE minute = 59 & hour < 23
    THEN (* \underline{minute := 1}*); hour := hour + 1 END;
  next_day =
    PRE minute = 59 & hour = 23
    THEN minute := 0 ; hour := 0 END END
\end{lstlisting}
\vspace{-3pt}
(b) Clock Model 2 (CM2)
\end{minipage}
\\
\begin{minipage}[t]{0.45\textwidth}
\centering
\begin{lstlisting}[frame=single,basicstyle=\scriptsize,columns=fullflexible]
MACHINE CM3 // Clock Model 3
VARIABLES hour, minute
INVARIANT hour : 0..23 & minute : 0..59
INITIALISATION hour := 0; minute := 0
OPERATIONS
  inc_minute = PRE (* \underline{minute $ < $ 58}*)
    THEN minute := minute + 1 END;
  (*\underline{inc\_minute\_part\_2 =}*) (*\underline{PRE minute = 58}*)
    (*\underline{THEN minute := 59 END;}*)
  inc_hour = PRE minute = 59 & hour < 23
    THEN minute := 0; hour := hour + 1 END;
  next_day = PRE minute = 59 & hour = 23
    THEN minute := 0; hour := 0 END
  END
\end{lstlisting}
\vspace{-3pt}
(c) Clock Model 3 (CM3)
\end{minipage}
~~
\vspace{2pt}
\begin{minipage}[t]{0.45\textwidth}
\centering
\begin{lstlisting}[frame=single,basicstyle=\scriptsize,columns=fullflexible]
MACHINE CM4 // Clock Model 4
VARIABLES hour, minute
INVARIANT hour : 0..23 & minute : 0..59
INITIALISATION hour := 0; minute := 0
OPERATIONS
  inc_minute = PRE (*\underline{minute $<=$ 59}*)
    THEN minute := minute + 1
    END;
  inc_hour =
    PRE minute = 59 & (*\underline{hour $<=$ 23}*)
    THEN minute := 0; hour := hour + 1
    END;
  next_day = PRE minute = 59 & hour = 23
    THEN minute := 0; hour := 0 END END
\end{lstlisting}
\vspace{-3pt}
(d) Clock Model 4 (CM4)
\end{minipage}
\\
\begin{minipage}[t]{0.45\textwidth}
\centering
\begin{lstlisting}[frame=single,basicstyle=\scriptsize,columns=fullflexible]
MACHINE CM5 // Clock Model 5
VARIABLES hour, minute
INVARIANT hour : 0..23 & minute : 0..59
INITIALISATION hour := 0; minute := 0
OPERATIONS
  inc_minute = (*\underline{SELECT hour = 3 \& minute = 0} *)
    (*\underline{THEN hour := 6; minute := 0} *)
    (*\underline{WHEN minute $ < $ 59}*)
      (*\underline{\& not(hour = 5 \& minute = 29)}*)
    THEN minute := minute + 1 END;
  inc_hour = PRE minute = 59 & hour < 23
    THEN minute := 0; hour := hour + 1 END;
  next_day = PRE minute = 59 & hour = 23
    THEN minute := 0; hour := 0 END END
\end{lstlisting}
\vspace{-3pt}
(e) Clock Model 5 (CM5)
\end{minipage}
~~
\begin{minipage}[t]{0.45\textwidth}
\centering
\begin{lstlisting}[frame=single,basicstyle=\scriptsize,columns=fullflexible]
MACHINE CM6 // Clock Model 6
VARIABLES hour, minute
INVARIANT hour : 0..23 & minute : 0..59
INITIALISATION hour := 0; minute := 0
OPERATIONS
  inc_minute = PRE minute < 59
    THEN minute := minute + 1 END;
  inc_hour = PRE minute = 59 & hour < 23
    THEN minute := 0; hour := hour + 1 END;
  next_day = PRE minute = 59 & hour = 23
    THEN minute := 0; hour := 0 END;
  (*\underline{set\_time =}*) (*\underline{ANY hh, mm}*)
    (*\underline{WHERE hh : 0..23 \& mm : 0..59}*)
    (*\underline{THEN hour := hh; minute := mm END}*) END
\end{lstlisting}
\vspace{-3pt}
(f) Clock Model 6 (CM6)
\end{minipage}
\caption{Clock models for explaining the criteria of B model quality.}
\label{fig:clock_models}
\end{figure*}

In our work, we use design models and state transition systems in the B method \cite{DBLP:books/daglib/0015096} as examples to develop criteria of design model quality. Fig. \ref{fig:clock_models} shows six B design models. In these models, initial states are derived from INITIALISATION clauses, and operations are specified under OPERATIONS clauses. Pre-conditions of operations are described using predicates between PRE (or WHERE) and THEN, and post-conditions are described using substitutions between THEN and END. Desired properties are described using predicates after INVARIANT clauses. These models will be used in Section \ref{sec:criteria} to explain our proposed criteria.

\section{Criteria for Measuring Quality of \\Formal Design Models}
\label{sec:criteria}

To evaluate the quality of B models, we derive a number of criteria from ISO/IEC 25010. Among the eight characteristics of ISO/IEC 25010, six characteristics including functional suitability, reliability, performance efficiency, usability, security and maintainability can be reflected by B models. The remaining two characteristics, i.e., compatibility and portability, cannot be reflected by B models because these characteristics are measured after software is actually implemented, installed on a real machine and used by users. In the following parts, we define a set of criteria for measuring the quality of B models based on the six characteristics.

Fig. \ref{fig:clock_models} shows examples of clock models to explain the proposed criteria. These models are expected to represent the change of minutes and hours within a day using two variables ``$ hour $" and ``$ minute $". All models have the same invariant that restricts the domain of the variables. The variable $ hour $ must be an integer between 0 and 23, and the variable $ minute $ must be an integer between 0 and 59. The two variables can represent time from 0:00 to 23:59. The initial values of $ hour $ and $ minute $ are 0, and their values are changed via ``inc\_minute", ``inc\_hour" and ``next\_day" operations. The operations in Clock Model 1 (CM1) are all correct. The inc\_minute operator can count $ minute $ from 0 to 59. When $ minute = 59 $, the inc\_hour operator increases $ hour $ by 1 and resets $ minute $ to 0, and the pre-condition $ hour < 23 $ ensures that the value of $ hour $ does not exceed 23 after running this operation. The operations in other models may be faulty and slightly different from their counterparts in CM1, and such differences are underlined in Fig. \ref{fig:clock_models}. These differences will be explained later when explaining the proposed criteria.

\subsection{Functional Suitability}

\textbf{Functional suitability} of B models can be measured by observing whether given B models provide operations that meet desired functional requirements specified by users. To describe the functional requirements, the users can directly provide a set of required state transitions $ T_{required} $, or provide a set of descriptions that can infer $ T_{required} $. For example, functional requirements for the clock models can be specified using the following state transitions:
\begin{itemize}
\item totally 1,416 state transitions of the form [($h$, $m$), inc\_minute, ($h$, $m+1$)] ($ h=0,\ldots,23 $ and $ m=0,\ldots,58 $), which represent the first 59 minutes within an hour, e.g., [(0, 0), inc\_minute, (0, 1)], [(9, 34), inc\_minute, (9, 35)] and [(23, 58), inc\_minute, (23, 59)].
\item totally 23 state transitions of the form [($h$, 59), inc\_hour, ($h+1$, 0)] ($ h=0,\ldots,22 $), which represent the last one minute within an hour except the last one minute of a day, e.g., [(0, 59), inc\_hour, (1, 0)], [(9, 59), inc\_hour, (10, 0)] and [(22,59), inc\_hour, (23, 0)].
\item a state transition [(23, 59), next\_day, (0, 0)], which represents the last one minute of a day.
\end{itemize}
There are totally 1,440 required state transitions for the clock models. If the clock models can exactly derive the required state transitions, then their functional suitability is high because they meet the functional requirements. Generally, suppose that $ T_{required} $ and a B model $ M $ is given, the sub-characteristics of functional suitability, i.e., functional completeness, functional correctness and functional appropriateness, can be measured using the following criteria.

The \textbf{functional completeness} of $ M $ can be reflected by observing whether the required state transitions can be derived by $ M $. Let $ T_{derived} $ denote a set containing all state transitions derived by $ M $, the \textbf{total functional completeness (TFComp)} of $ M $ can be computed via
\begin{equation}
TFComp = |T_{derived} \cap T_{required}| ~/~ |T_{required}|
\end{equation}
In this formula, $ T_{derived} \cap T_{required} $ collects derived state transitions that are required by the users. The whole formula computes the percentage of the derived state transitions with respect to all the required state transitions. For example, the total functional completeness of CM1 is 1 because CM1 can derive all the required state transitions. Regarding Clock Model 2 (CM2) in Fig. \ref{fig:clock_models} (b), the underlined substitution ``minute := 1" results in 23 unexpected state transitions of the form [($h$, 59), inc\_hour, ($h+1$, 1)] ($ h=0,\ldots,22 $), while 23 required state transitions of the form [($h$, 59), inc\_hour, ($h+1$, $0$)] ($ h=0,\ldots,22 $) and other 23 required state transitions of the form [($h$, 0), inc\_minute, ($h$, 1)] ($ h=1,\ldots,23 $) are missing. As $ T_{derived} \cap T_{required} $ includes $ 1,440 - 23 - 23 = 1,394 $ state transitions, the total functional completeness of CM2 is $ 1,394 / 1,440  \approx 0.968 $.

Although CM2 derives the 23 unexpected state transitions, these transitions can partially reach the functional requirements. For example, [(1, 59), inc\_hour, (2, 1)], which is an unexpected state transition, is very similar to the required state transition [(1, 59), inc\_hour, (2, 0)]. In this case, minor corrections to the given B model can make it fit the functional requirements, so that its functional completeness can be increased. To measure the possibility of turning unexpected state transitions to required state transitions, \textbf{partial functional completeness (PFComp)} is introduced:
\begin{equation}
PFComp = \cfrac{Similarity(T_{derived},T_{required})}{Size(T_{required})}
\end{equation}
In this formula, $ Similarity(T_1,T_2) $ computes the similarity between two sets of state transitions $ T_1 $ and $ T_2 $ by finding maximum alignments from state transitions in $ T_1 $ to state transitions in $ T_2 $. Let $ [(v_1,\ldots,v_N),\alpha,(v'_1,\ldots,v'_N)] $ and $ [(w_1,\ldots,w_N),\beta,(w'_1,\ldots,w'_N)] $ denote two state transitions, they can be flattened to be two lists $ [v_1,\ldots,v_N,\alpha,v'_1,\ldots,v'_N] $ and $ [w_1,\ldots,w_N,\beta,w'_1,\ldots,w'_N] $. The elements in the first lists are compared with their counterparts in the second list, so that the number of equal elements between the two lists can be counted. For example, the two state transitions [(1, 59), inc\_hour, (2, 1)] and [(1, 59), inc\_hour, (2, 0)] can be flattened to be [1, 59, inc\_hour, 2, 1] and [1, 59, inc\_hour, 2, 0], and the number of equal elements between them is 4. To find maximum alignments between the two sets of state transitions $ T_1 $ and $ T_2 $, the number of equal elements between each state transition in $ T_1 $ and each state transition in $ T_2 $ is counted. When the number of equal elements reaches a highest value, the corresponding two state transitions are aligned. Note that a state transition in $ T_1 $ can be aligned with at most one state transition in $ T_2 $, and vice versa. After finding all aligned state transitions, the total number of aligned elements is returned to be the similarity between $ T_1 $ and $ T_2 $. For example, as mentioned before, CM2 has 23 derived state transitions of the form [($h$, 59), inc\_hour, ($h+1$, 1)] ($ h=0,\ldots,22 $), which can be aligned with the 23 required state transitions of the form [($h$, 59), inc\_hour, ($h+1$, $0$)] ($ h=0,\ldots,22 $), and the number of equal elements in each alignment is 4 out of 5. Besides, CM2 has 1,394 derived state transitions that are exactly the same as the required state transitions. As a result, the value of $ Similarity(T_{derived},T_{required}) $ for CM2 is $ 4 \times 23 + 5 \times 1,394 = 7,062 $. $ Size(T) $ is the size of a set of state transitions $ T $. It is computed by flattening all state transitions in $ T $ to be lists and counting the total number of elements in the lists. For example, the clock models require 1,440 state transitions with 5 elements, i.e., two elements in pre-states, an operation identifier and two elements in post-states, so that the value of $ Size(T_{required}) $ for the clock models is $ 5 \times 1,440 = 7,200 $. The partial functional completeness of CM2 is $ 7,062 / 7,200 \approx 0.981 $.

The \textbf{functional correctness} of $ M $ can be measured by observing whether all the state transitions in $ T_{derived} $ are precisely derived. To measure the precision of $ T_{derived} $, \textbf{total functional correctness (TFCorr)} of $ M $ is computed via
\begin{equation}
TFCorr = |T_{derived} \cap T_{required}| ~/~ |T_{derived}|
\end{equation}
This formula computes the percentage of correctly derived state transitions with respect to all the derived state transitions. Recall the example of CM2, $ T_{derived}~\cap~T_{required} $ has 1,394 state transitions, and $ T_{derived} $ has $ 1,394 + 23 = 1,417 $ state transitions. As a result, the total functional correctness of CM2 is $ 1,394 / 1,417 \approx 0.984 $. Regarding CM1, its total functional correctness is 1 because CM1 derives all the required state transitions and does not derive any unnecessary state transitions.

As discussed before, CM2 derives the 23 incorrect state transitions of the form [($h$, 59), inc\_hour, ($h+1$, 1)] ($ h=0,\ldots,22 $). Minor corrections that change these state transitions to [($h$, 59), inc\_hour, ($h+1$, 0)] ($ h=0,\ldots,22 $) can make CM2 precisely fit the functional requirements, so that the functional correctness can be increased to 1. To measure the possibility of turning incorrect state transitions to be correct, \textbf{partial functional correctness (PFCorr)} is introduced:
\begin{equation}
PFCorr = \cfrac{Similarity(T_{derived},T_{required})}{Size(T_{derived})}
\end{equation}
For example, $ Similarity(T_{derived},T_{required}) $ for CM2 is $ 7,062 $, and $ Size(T_{derived}) $ for CM2 is $ 5 \times 1,417 = 7,085 $, so that the partial functional correctness is $ 7,062 / 7,085 \approx 0.997 $.

As both the functional correctness and the functional completeness of CM2 are lower than 1, CM2 is considered incomplete and inconsistent with respect to the functional requirements. It is possible that a B model is complete and inconsistent, or incomplete and consistent. If $ T_{required} \subsetneq T_{derived} $, then the model is complete and inconsistent because the model derives not only all required state transitions, but also a number of unexpected state transitions. If $ T_{derived} \subsetneq T_{required} $, then the model is incomplete and consistent because the model does not derive any unexpected state transitions, but a set of required state transitions are missing.

The \textbf{functional appropriateness} of $ M $ can be measured by observing whether all required pairs of pre- and post-states (abbreviated as ``pairs" in later discussions) are achievable regardless of operations. Let $ P_{derived} $ be a set containing all pairs of the form $ (s,t) $ such that $ (s,\alpha,t) \in T_{derived} $, and let $ P_{required} $ be a set containing all pairs of the form $ (s,t) $ such that $ (s,\alpha,t) \in T_{required} $, \textbf{total functional appropriateness (TFAppr)} of $ M $ is computed via
\begin{equation}
TFAppr = |P_{derived} \cap P_{required}| ~/~ |P_{required}|
\end{equation}
In this formula, $ P_{derived} \cap P_{required} $ collects derived pairs that are required by the users. The whole formula computes the percentage of the collected pairs with respect to all the required pairs. Intuitively, if a required state transition is not achievable using the required operation, but is achievable using an alternative operation, then the alternative operation is appropriate with respect to the functional requirements. For example, in Fig. \ref{fig:clock_models} (c), Clock Model 3 (CM3) has two operations inc\_minute and inc\_minute\_part\_2 that derive state transitions of the form [($h$, $m$), inc\_minute, ($h$, $m+1$)] ($ h=0,\ldots,23 $ and $ m=0,\ldots,57 $) and [($h$, 58), inc\_minute\_part\_2, ($h$, 59)] ($ h=0,\ldots,23 $). The above state transitions lead to pairs of the form [($h$, $m$), ($h$, $m+1$)] ($ h=0,\ldots,23 $ and $ m=0,\ldots,58 $), and these pairs are appropriate with respect to the required state transitions of inc\_minute, which are of the form [($h$, $m$), inc\_minute, ($h$, $m+1$)] ($ h=0,\ldots,23 $ and $ m=0,\ldots,58 $). This means that the functional requirements of inc\_minute are achievable by integrating inc\_minute and inc\_minute\_part\_2 in CM3. As a result, the total functional appropriateness of CM3 is 1. Regarding CM2, its $ P_{derived} $ contains 23 pairs of the form [($h$, 59), ($h$, 1)] ($ h=0,\ldots,22 $) that are inappropriate with respect to the required [($h$, 59), ($h$, 0)] ($ h=0,\ldots,22 $). Moreover, CM2 fails to derive 23 required state transitions of the form [($h$, 0), inc\_minute, ($h$, 1)] ($ h=1,\ldots,23 $), so that 23 pairs of the form [($h$, 0), ($h$, 1)] ($ h=1,\ldots,23 $) are missing. As the number of required pairs are 1,440, the number of appropriate pairs is $ 1,440 - 23 - 23 = 1,394$. The total functional appropriateness of CM2 is $ 1,394 / 1,440 \approx 0.968 $.

It is possible that minor changes can turn inappropriate pairs to be appropriate. For example, in CM2, [(2, 59), inc\_hour, (3, 1)] is inappropriate, but replacing 1 with 0 can make it be an appropriate state transition [(2, 59), inc\_hour, (3, 0)]. To measure the possibility that minor changes can turn inappropriate state transitions to be appropriate, \textbf{partial functional appropriateness (PFAppr)} is introduced:
\begin{equation}
PFAppr = \cfrac{Similarity(P_{derived},P_{required})}{Size(P_{required})}
\end{equation}
For example, CM2 has 23 inappropriate pairs of the form [($h$, 59), ($h$, 1)] ($ h=0,\ldots,22 $) that can become appropriate by replacing 1 with 0, and the other 1,394 pairs are all appropriate. As the number of variables in each pair is 4, the value of $ Similarity(P_{derived},P_{required}) $ is $ 4 \times 1,394 + (4-1) \times 23 = 5,645 $. The value of $ Size(P_{required}) $ is $ 4 \times 1,440 = 5,760 $. As a result, the partial functional appropriateness is $ 5,645 / 5,760 \approx 0.980 $.

\subsection{Security and Reliability}

In B models, properties of \textbf{security} and \textbf{reliability} are usually specified using invariants. As assertions in B models are a special type of invariants, in the following discussions, ``invariants" include assertions, and ``invariant violations" include assertion violations. In our work, deadlock-freeness, which requires that each state has at least one outgoing transition, is considered as an inherent invariant and is always checked even if it is not specified in the B models.

Sub-characteristics of security include confidentiality, integrity, non-repudiation, accountability and authenticity, and sub-characteristics of reliability include maturity, availability, fault tolerance and recoverability. Among these sub-characteristics, properties of confidentiality, integrity, authenticity and maturity can be specified using invariants. \textbf{Confidentiality} requires that any unauthorised accounts do not access protected data. To ensure the confidentiality, an invariant can be specified to check that all derived state transitions do not lead to any states where an unauthorised account is reading a protected file. \textbf{Integrity} requires that any unauthorised accounts do not access and modify protected data. To ensure the integrity, an additional restriction can be add into the invariant of confidentiality to check that the unauthorised account is not writing to the protected file. \textbf{Authenticity} requires that an accessed identity is the corresponding claimed identity. To ensure the authenticity, an invariant can be specified to check if the claimed identity and the accessed identity remain the same at any time after accessing the identity and before releasing the identity. \textbf{Maturity} requires that a model performs required functions under normal operations and can be measured by observing: (1) whether the model meets the functional requirements, which can be measured using the criteria of the functional completeness and the functional consistency, and (2) whether operations in the model can normally run without violating any invariants, which can be measured together with confidentiality, integrity and authenticity using \textbf{invariant satisfability}:
\begin{equation}
Invariant~Satisfability = |T^{\top}_{derived}| ~/~ |T_{derived}|
\label{eq:invsat}
\end{equation}
where $ T^{\top}_{derived} $ is a set containing all derived state transitions that do not trigger any invariant violations. This formula computes the percentage of correctly derived state transitions with respect to all derived state transitions. For example, in Fig. \ref{fig:clock_models} (d), Clock Model 4 (CM4) violates its invariants because inc\_minutes can be activated at the end of the 59th minute, resulting in 24 faulty state transitions of the form [($h$, 59), inc\_minute, ($h$, 60)] ($ h=0,\ldots,23 $) that violate the invariant ``minute : 0..59". Moreover, inc\_hour can be activated at the end of 23:59, resulting in a faulty state transition [(23, 59), inc\_hour, (24, 0)]. Besides, CM4 can derive the 1,440 correct state transitions in $ T_{required} $. As a result, $ T^{\top}_{derived} $ contains 1,440 state transitions, $ T_{derived} $ contains $ 1,440 + 24 + 1 = 1,465 $ state transitions, and the invariant satisfability of CM4 is $ 1,440 / 1,465 \approx 0.983 $.

\textbf{Availability} and \textbf{non-repudiation} can be measured by observing whether operations are accessible without invariant violations. Let $ F_{required} $ denote a set containing all required operations, which derives the state transitions in $ T_{required} $, and let $ F_{derived}^{\top} $ denote a set containing all accessible and operations without any invariant violations, which derives the state transitions in $ T^{\top}_{derived} $, the following formula is used to compute availability and non-repudiation:
\begin{equation}
Availability = |F_{derived}^{\top} \cap F_{required}| ~/~ |F_{required}|
\end{equation}
This formula computes the percentage of the correctly accessed operations with respect to the required operations. For example, CM4 triggers invariant violations because inc\_minutes and inc\_hour derive faulty state transitions. As $ F_{required} $ contains three operations, i.e., inc\_minutes, inc\_hour and next\_day, and only next\_day does not trigger any invariant violations, the availability of CM4 is $ 1 / 3 \approx 0.333 $.

\textbf{Accountability} can be measured by observing whether the derived states are uniquely traceable. The significance of uniquely traceable states is that if such a state triggers an invariant violation, the cause of the invariant violation can be found by tracing back from this state to an initial state. A state $ s $ is uniquely traceable if there exists at most one state transition $ [s_{pre},\alpha,s] \in T_{derived} $ such that $ s_{pre} $ is an initial state, or a uniquely traceable state. Suppose that $ S^1_{traceable} $ is a set containing all states that are in $ S_{derived} $ and with at most one ingoing state transition in $ T_{derived} $, the accountability is:
\begin{equation}
Accountability = |S^1_{traceable}| ~/~ |S_{derived}|
\end{equation}
This formula computes the percentage of states with at most one ingoing transition with respect to all the derived states. For example, as CM1 only derives the 1,440 states representing each minute from 0:00 to 23:59, each state can only be traced back to exactly one state representing the previous one minute. As a result, the accountability of CM1 is 1. On the other hand, If a model derives four state transitions such as [(0, 0), inc\_x,(1, 0)], [(0, 0), inc\_y,(0, 1)], [(0, 1), inc\_x,(1, 1)] and [(1, 0), inc\_y,(1, 1)], then (1, 1) can be traced back to (0, 1) and (1, 0). As a result, $ S^1_{traceable} $ contains three states (0, 0), (0, 1) and (1, 0), and $ S_{derived} $ contains four states (0, 0), (0, 1), (1, 0) and (1, 1), so that the accountability is $ 3 / 4 $.

\textbf{Fault tolerance} can be measured by observing whether operations in the model $ M $ perform their intended functions when faults occur. Before observation, two types of imitated faults are randomly injected into $ M $. The first type of faults is a set $ T_{extra} $ of extra state transitions that are inserted into $ M $. The second type of faults is a set $ T_{missing} $ of missing state transitions that are removed from $ M $. The state transitions in $ T_{extra} $ and $ T_{missing} $ are randomly generated. After applying the changes to $ M $, the resulting model $ M_{changed} $ is checked using the ProB model checker to obtain a set of state transitions $ T_{changed} $. To observe the impact of the changes to the other state transitions, the changes themselves should be masked before observation, so that a set of state transitions $ U_{changed} = (T_{changed} \cup T_{missing}) - T_{extra} $ is produced. After producing $ U_{changed} $, it is checked against the invariants to obtain a set $ U^{\top}_{changed} $ containing all state transitions that satisfy the invariants and a set $ U^{\bot}_{changed} $ containing all state transitions that violate the invariants. Based on the above results, we can measure the impact of the imitated faults via:
\begin{equation}
Fault~Tolerance = 1 - |U^{\bot}_{changed}| ~/~ |U_{changed}|
\end{equation}
For example, inserting [(3, 0), inc\_minute, (12, 0)] to and removing [(5, 29), inc\_minute, (5, 30)] from CM1 lead to Clock Model 5 (CM5) in Fig. \ref{fig:clock_models} (e). If $ M_{changed} $ is CM5, $ T_{changed} $ will include state transitions representing each minute from 0:00 to 5:29 and from 12:00 to 23:59 and include [(3, 0), inc\_minute, (12, 0)]. In this case, $ U_{changed} $ includes 1,050 state transitions representing each minute from 0:00 to 5:30 and from 12:00 to 23:59, which means that the two changes result in a gap between 5:30 and 12:00. This gap makes the state $ (5,30) $ have no outgoing transitions in $ U_{changed} $, so that $ U^{\bot}_{changed} $ only contains $ [(5, 29), inc\_minute, (5, 30)] $, and $ U^{\top}_{changed} $ contains the remaining 1,049 state transitions. As a result, the fault tolerance is $ 1 - 1 / 1,050 \approx 0.999 $. It is very important to note that we only apply the two changes to the model to simplify the explanation. When actually evaluating $ M $, $ M_{changed} $ is produced by applying a sufficient number of changes to $ M $.

Moreover, the fault tolerance requires $ M $ to perform its intended functions when faults occur, which means that $ M $ is expected to re-establish correct state transitions as far as possible. The possibility of re-establishment can be measured together with \textbf{recoverability} via:
\begin{equation}
Recoverability = |U^{\top}_{changed} \cap T_{derived}| ~/~ |T_{derived}|
\end{equation}
This formula computes the percentage of re-establishable state transitions with respect to intended state transitions. For example, in CM5, the removal of [(5, 29), inc\_minute, (5, 30)] blocks the derivation of the intended state transitions from (5, 30), and the insertion of $ [(3, 0), inc\_minute, (12, 0)] $ enables the re-establishment of the intended state transitions from (12, 0). In this case, $ U^{\top}_{changed} \cap T_{derived} $ contains all the 1,049 state transitions in $ U^{\top}_{changed} $, so that the recoverability is $ 1,049 / 1,440 \approx 0.728 $. An effective way to improve the recoverability is to define a new operation $ set\_time $ that allows the users to set the time to any points from 0:00 to 23:59. Clock Model 6 (CM6) in Fig. \ref{fig:clock_models} (f) shows the code of $ set\_time $, which increases the recoverability to 1.

\subsection{Maintainability}
\label{sec:maintainability}

\textbf{Maintainability} measures the possibility of modifying the given B model $ M $. Analysability and modifiability, which are sub-characteristics of maintainability, are computed based on $ M_{changed} $ because they are observed after changing $ M $. The \textbf{analysability} measures whether changes to $ M $ can significantly impact its functionality and the satisfiability to the invariants. Intuitively, a maintainer who is not developers of a given model may consider the model as a black-box. If a slight change to the model can significantly change the behaviour of the model, it will be relatively easy for the maintainer to decide whether to and how to modify the component. For example, removing [(5, 29), inc\_minute, (5, 30)] from CM1 blocks the derivation of the state transitions after (5, 30), which makes the model miss a significantly number of state transitions. Consequently, the maintainer may infer that inc\_minute plays a crucial role in $ M $, so that the modification of existing state transitions with inc\_minute should be avoided. To quantify the analysability, we suggest functional analysability and fault analysability.

\textbf{Functional analysability} measures whether changes to $ M $ can significantly impact its functionality and is computed via:
\begin{equation}
Functional~Analysability = 1 - \cfrac{|T_{derived} \cap U_{changed}|}{|T_{derived} \cup U_{changed}|}
\label{eq:functional_analysability}
\end{equation}
In this function, if the functionality of $ M $ is significantly impacted by the changes, $ U_{changed} $ will be significantly different from $ T_{derived} $. In this case, the functional analysability is high because their common part $ T_{derived} \cap U_{changed} $ only has a few elements. On the other hand, if the changes does not cause significant differences between $ U_{changed} $ and $ T_{derived} $, the functional analysability will be low. For example, after changing CM1 to CM5, $ T_{derived} \cap U_{changed} $ contains the $ 1,050 $ state transitions in $ U_{changed} $, and $ T_{derived} \cup U_{changed} $ contains the $ 1,440 $ state transitions in $ T_{derived} $, so that the functional analysability is $ 1 - 1,050 / 1,440 \approx 0.271 $. Again, please note that we only apply two changes to CM1 to simplify the explanation. A considerable number of changes should be applied to $ M $ when actually evaluating the analysability of $ M $.

\textbf{Fault analysability} measures whether changes to $ M $ can significantly impact its satisfiability to the invariants:
\begin{equation}
Fault~Analysability = 1 - \cfrac{|T^{\bot}_{derived} \cap U^{\bot}_{changed}|}{|T^{\bot}_{derived} \cup U^{\bot}_{changed}|}
\label{eq:fault_analysability}
\end{equation}
In this function, if the changes to $ M $ can newly introduce a significantly number of invariant violations or eliminate a significantly number of invariant violations, $ U^{\bot}_{changed} $ will be significantly different from $ T^{\bot}_{derived} $. As $ T^{\bot}_{derived} \cap U^{\bot}_{changed} $ only has a few elements, the fault analysability is high. On the other hand, if the changes does not cause significant differences on the satisfiability to the invariants, the fault analysability will be low. For example, after changing CM1 to CM5, $ T^{\bot}_{derived} $ is empty, and $ U^{\bot}_{changed} $ contains only one state transition [(5, 29), inc\_minute, (5, 30)] because (5, 30) has no outgoing transitions in $ U_{changed} $. As $ T^{\bot}_{derived} \cap U^{\bot}_{changed} $ is empty, the fault analysability is $ 1 $.

\textbf{Modifiability} measures whether $ M $ can be modified without introducing invariant violations and significantly influencing its existing quality. As modifiability can be measured by reusing criteria for other (sub-)characteristics such as analysability, recoverability, modularity and learnability, we do not define new criteria for modifiability.

\textbf{Modularity} measures whether $ M $ has discrete operations such that a change to an operation $ \alpha $ has minimal impact on the other operations. To observe the impact of $ \alpha $ on the other operations, a model $ M_{\Delta(\alpha)} $ is produced by randomly inserting new state transitions with $ \alpha $ into or removing existing state transitions with $ \alpha $ from $ M $. Next, state transitions of $ M_{\Delta(\alpha)} $ are derived using the ProB model checker, and all derived state transitions not derived by $ \alpha $ are collected into a set $ T^{\neg \alpha}_{\Delta(\alpha)} $. After that, all state transitions derived by $ M $ and not derived by $ \alpha $ are collected into a set $ T^{\neg \alpha}_{derived} $. The \textbf{modularity} of $ \alpha $ is computed via:
\begin{equation}
Modularity(\alpha) = \cfrac{|T^{\neg \alpha}_{derived} \cap T^{\neg \alpha}_{\Delta(\alpha)}|}{|T^{\neg \alpha}_{derived} \cup T^{\neg \alpha}_{\Delta(\alpha)}|}
\end{equation}
Intuitively, if $ \alpha $ is an individual module that has minimum impact on other operations, the changes to $ \alpha $ will have minimum impact on state transitions not derived by $ \alpha $. Consequently, the difference between $ T^{\neg \alpha}_{\Delta(\alpha)} $ and $ T^{\neg \alpha}_{derived} $ is minimum, so that $ Modularity(\alpha) $ will be high. On the other hand, if the impact is significant, $ Modularity(\alpha) $ will be low because $ T^{\neg \alpha}_{\Delta(\alpha)} $ are significantly different from $ T^{\neg \alpha}_{derived} $. CM5 can be an example of $ M_{\Delta(\alpha)} $ with respect to inc\_minute in CM1. In this example, $ T^{\neg \alpha}_{derived} $ contains 24 elements, i.e., the 23 state transitions derived by inc\_hour and the only one state transition derived by next\_day, while $ T^{\neg \alpha}_{\Delta(\alpha)} $ contains all the state transitions in $ T^{\neg \alpha}_{derived} $ except $ [(5, 59), inc\_hour, (6, 0)] $. As a result, the modularity is $ (24 - 1) / 24 \approx 0.958 $. It is important to note that we only apply two changes to CM1 to produce CM5 to simply the explanation. When actually evaluating the modularity of $ \alpha $, a sufficient number of changes should be applied to $ M $ to produce $ M_{\Delta(\alpha)} $. After obtaining the modularity values of all operations, the weighted sum of these modularity values is considered as the modularity of the whole model. The weight of $ \alpha $ is the proportion of state transitions with $ \alpha $ with respect to all the derived state transitions.

\textbf{Reusability} of $ M $ is measured by observing whether each operation can be used in different situations:
\begin{equation}
Reusability = 1 - |F_{derived}| ~/~ |T_{derived}|
\end{equation}
where $ F_{derived} $ is a set containing all operations in the derived state transitions. For example, as CM1 has 3 operations and derives 1,440 operations, the reusability of CM1 is $ 1 - 3 / 1,440 = 0.998 $. Besides, \textbf{testability} is measured by observing model checking CPU time within a given time limit, which is introduced in Section \ref{sec:performance}.

\subsection{Performance Efficiency and Usability}
\label{sec:performance}

In this section, we propose criteria for the remaining two characteristics, i.e., performance efficiency and usability. \textbf{Performance efficiency} of the B model $ M $ can be measured by observing the usage of time and hardware resources during model checking. Sub-characteristics of performance efficiency includes time behaviour, resource utilisation and capacity. The \textbf{time behaviour} can be measured by observing \textbf{model checking CPU time} $ T^{CPU}_{MC} $. The \textbf{resource utilisation} can be measured by observing \textbf{peak memory usage}. The \textbf{capacity} is calculated via
\begin{equation}
Capacity = |S_{derived}| + |T_{derived}|
\label{eq:capacity}
\end{equation}
where $ S_{derived} $ is a set containing all derived states. For example, as CM1 can display time from 0:00 to 23:59, $ S_{derived} $ has totally 1,440 states. Moreover, $ T_{derived} $ has totally 1,440 state transitions. Thus, the capacity of CM1 is $ 2,880 $.

\textbf{Useability} is measured by observing whether $ M $ can efficiently and effectively achieve desired goals. Except appropriateness recognizability, user error protection and learnability, the other sub-characteristics of useability need to be measured based on actual software products, but not based on B models. Therefore, we will only provide criteria of appropriateness recognizability, user error protection and learnability. \textbf{Appropriateness recognisability} can be measured by (1) functional appropriateness, which has been described before, and (2) \textbf{goal appropriateness (GAppr)}:
\begin{equation}
GAppr = |G^{\top}| ~/~ |G|
\label{eq:goal_appropriateness}
\end{equation}
where $ G $ is a set of goal predicates, and $ G^{\top} $ is a set containing all goal predicates that can be achieved by $ M $. For example, if two goals for CM1 are ``G1: hour + minute $<$ 10" and ``G2: hour $>$ 26 \& minute $<$ 10", then G1 is achievable, while G2 is not. As a result, the goal appropriateness of CM1 is 0.5. \textbf{Learnability} can be measured by counting the number of words in the source code of $ M $. $ M $ includes specifications that may be used to produce instructions of final software products. If $ M $ itself is easy to be understood, the instructions will be more learnable. The learnability is computed via
\begin{equation}
Learnability = 1 - Min(N_{words},N^{limit}_{words}) ~/~ N^{limit}_{words}
\label{eq:learnability}
\end{equation}
where $ N_{words} $ is the number of words in $ M $, and $ N^{limit}_{words} $ is a word limit given by the users. Besides, \textbf{User error protection} concerns whether $ M $ has mechanisms to avoid the users' faults and can be measured using the criteria of security and reliability.

\section{Evaluation}
\label{sec:evaluation}

The proposed criteria were implemented as a B model quality evaluator with the ProB model checker \cite{DBLP:journals/sttt/LeuschelB08}. To test the evaluator, we conducted an experiment on a server equipped with two Intel Xeon Gold 6130 Processor, 256GB DDR4 RAM and the Ubuntu Server 18.04 operating system. The proposed criteria were used to compute the quality of the Volvo cruise controller (VCC) model, which was downloaded from the ProB Public Example Repository. Based on the original VCC model (VCC0), we produced an adaptation (VCC1) by (1) inserting new state transitions, (2) modifying existing state transitions, and (3) deleting existing state transitions. Moreover, state transitions of VCC0 were extracted to be the required state transitions, and a set of goals were randomly generated.\footnote{The dataset is available on \url{https://github.com/cchrewrite/VCC-data}.} All the proposed criteria were used to measure the quality of VCC0 and VCC1.

\begin{table}[htbp]
\caption{The Quality of Volvo Cruise Controller Models}
\begin{center}
\begin{tabular}{|c|c|c|c|c|}
\hline
\multirow{2}{*}{\textbf{Model}}& \multicolumn{4}{|c|}{\textbf{Quality Measurement}} \\
\cline{2-5}
& \textbf{\textit{TFComp}} & \textbf{\textit{PFComp}} & \textbf{\textit{TFCorr}} & \textbf{\textit{PFCorr}} \\
\hline
VCC0 & 1.000 & 1.000 & 1.000 & 1.000 \\
\hline
VCC1 & 0.913 & 0.997 & 0.770 & 0.842 \\
\hline
\hline
& \textbf{\textit{TFAppr}} & \textbf{\textit{PFAppr}} & \textbf{\textit{Inv. Sat.}} & \textbf{\textit{Availability}} \\
\hline
VCC0 & 1.000 & 1.000 & 1.000 & 1.000 \\
\hline
VCC1 & 0.887 & 0.996 & 0.800 & 0.846 \\
\hline
\hline
& \textbf{\textit{Accountability}} & \textbf{\textit{Fau. Tol.}} & \textbf{\textit{Recoverability}} & \textbf{\textit{Fun. Ana.}} \\
\hline
VCC0 & 0.000 & 0.352 & 0.089 & 0.878 \\
\hline
VCC1 & 0.328 & 0.234 & 0.065 & 0.836 \\
\hline
\hline
& \textbf{\textit{Fau. Ana.}} & \textbf{\textit{Modularity}} & \textbf{\textit{Reusability}} & \textbf{\textit{CPU Time}} \\
\hline
VCC0 & 0.900 & 0.913 & 0.983 & 6.228 (s) \\
\hline
VCC1 & 0.841 & 0.746 & 0.986 & 7.982 (s) \\
\hline
\hline
& \textbf{\textit{Peak Mem.}} & \textbf{\textit{Capacity}} & \textbf{\textit{GAppr}} & \textbf{\textit{Learnability}}\\
\hline
VCC0 & 0.166 (GB) & 27,052 & 1.000 & 0.982 \\
\hline
VCC1 & 0.168 (GB) & 32,500 & 0.890 & 0.982 \\
\hline
\end{tabular}
\label{tab:eval_vcc}
\end{center}
\end{table}

Table \ref{tab:eval_vcc} shows the divergence of quality between VCC0 and VCC1. Comparing to VCC0, VCC1 had lower TFComp, PFComp, TFAppr and PFAppr because a number of required state transitions were missing. A side-effect of the missing state transitions was the reduction of fault tolerance, recoverability and modularity because a number of paths for VCC1 to perform expected functions and recover from unexpected states were missing. Another side-effect was the reduction of GAppr because a number of paths to the goals were blocked. Moreover, VCC1 had lower TFCorr and PFCorr because it had a set of unexpected state transitions. Besides, these transitions might violate the invariants, reducing the invariant satisfiability and availability. Additionally, VCC1 had higher accountability, CPU time and capacity because it had a set of new state transitions and traceable states. The above results revealed that the proposed criteria could reflect the divergence of model quality from different aspects.

\section{Related Work}
\label{sec:related_work}

The proposed criteria are similar to a number of existing criteria of design model quality. In \cite{DBLP:journals/sqj/NelsonM07}, the semantic quality is used to measure whether a design model can precisely reflect knowledge in real world domains. As both the semantic quality and the functional suitability in ISO/IEC 25010 \cite{ISOIEC25010} are defined for correctness and completeness, the proposed TFComp, PFComp, TFCorr and PFCorr can be used to quantify the semantic quality. In \cite{DBLP:journals/tse/GouesNFW12}, the size of statements and the lines of code have been used to measure the readability of code, which can be supplemented using Eq. (\ref{eq:learnability}) for learnability. In automatic model construction, model quality can be measured using fitness functions that reflect the distance between state transition systems and observations in the real world \cite{DBLP:conf/gecco/BuzhinskyCUT14}. The fitness functions are similar to PFComp and PFCorr because they concern whether a model can fit functional requirements.

In \cite{DBLP:journals/infsof/MohagheghiDN09}, a wide range of criteria for measuring the quality of design models have been reviewed and classified into six categories: correctness, completeness, consistency, comprehensibility, confinement and changeability, which are named ``6C goals". In the 6C goals, correctness and completeness have similar definitions with their counterparts in ISO/IEC 25010 \cite{ISOIEC25010}, so that they can be measured using the proposed TFComp, PFComp, TFCorr and PFCorr. Consistency requires that a model does not contradicts with desired properties, and it can be measured using Eq. (\ref{eq:invsat}) for invariant satisfability and Eq. (\ref{eq:goal_appropriateness}) for goal appropriateness. Comprehensibility requires that a model is analysable to human users or tools, which can be measured using Eq. (\ref{eq:functional_analysability}) for functional analysability and Eq. (\ref{eq:fault_analysability}) for fault analysability. Confinement requires that a model should be precise and abstract. The precision of a model can be measured using TFCorr and PFCorr, but whether the model is abstract cannot be reflected by the proposed criteria. Changeability requires that a model can be easily improved by developers, which can be measured using the equations for maintainability in Section \ref{sec:maintainability}. Because of the above reasons, the proposed criteria can describe most aspects of the 6C goals.

\section{Conclusion}
\label{sec:conclusion}

Based on ISO/IEC 25010, we have proposed a set of criteria for evaluating design model quality. These criteria can be used together with model checking techniques to quantify the quality of state transition systems of design models. Moreover, the criteria have been implemented as a B model evaluator, and the experimental results on the Volvo cruise controller models have revealed that the evaluator can show the divergence of model quality between two different models.

In the future, we will use syntactic analysis methods to evaluate design models at the language level. Moreover, we will extend the proposed criteria to temporal logics. Additionally, the criteria will be extended to the field of automatic model repair.

\bibliographystyle{IEEEtran}
\bibliography{mybibfile}

\end{document}